\documentstyle[preprint,aps,floats]{revtex}
\input epsf.tex
\def\DESepsf(#1 width #2){\epsfxsize=#2 \epsfbox{#1}}
\def\Dbar{{\bar D}}

\def\prd#1#2#3{Phys. Rev. D {\bf #1}, #3 (#2)}

\begin{document}
\preprint{\vbox{\hbox{}}} \draft
\title {Hadronic decays of $B$ involving a tensor meson \\
through a $b \to c$ transition }
\author{C. S. Kim\footnote{cskim@mail.yonsei.ac.kr},
Jong-Phil Lee\footnote{jplee@phya.yonsei.ac.kr}, and Sechul
Oh\footnote{scoh@post.kek.jp}}
\address{Department of Physics and IPAP, Yonsei University, Seoul,
120-749, Korea}
\maketitle
\begin{abstract}
\noindent We re-analyze hadronic decays of $B$ mesons to a pseudoscalar ($P$)
and a tensor meson ($T$), or a vector meson ($V$) and a tensor meson,
through a $b \to c$ transition.  We discuss possible large uncertainties to
branching ratios (BR's) of the relevant modes, mainly arising from uncertainties
to the hadronic form factors for the $B \to T$ transition.
The BR's and CP asymmetries for $B \to PT$ and $VT$ decays are then calculated
by using the form factors given in the ISGW2 model (the improved version of
the original Isgur-Scora-Grinstein-Wise (ISGW) model).
We find that the estimated BR's of many modes are increased
by an order of magnitude, compared to the previous results calculated
within the ISGW model.
\end {abstract}

%%%%%%%%%%%%%%%%%%%%%%%%%%%%%%%%%%%%%%%%%%%%%%%%%%%%%%%%%%
%%%%%%%%%%%%%%%%%%%%%%%%%%%%%%%%%%%%%%%%%%%%%%%%%%%%%%%%%%
\newpage
\section{Introduction}

In the next few years plenty of new experimental data on rare decays of $B$ mesons
will be available from $B$ factory experiments
such as Belle, BaBar, BTeV, LHC-B and so on.
Experimentally several tensor mesons have been observed \cite{pdg},
such as the isovector $a_2$(1320), the isoscalars $f_2$(1270),
$f_2^{\prime}$(1525), $f_2$(2010), $f_2$(2300), $f_2$(2340),
$\chi_{c2}(1P)$, $\chi_{b2}(1P)$ and $\chi_{c2}(2P)$, and the
isospinors $K_2^*$(1430) and $D_2^*$(2460).  Experimental data on
the branching ratios (BR's) for $B$ decays involving a tensor meson ($T$)
in the final state provide only upper bounds\cite{pdg}: for instance,
for a $b \to c$ transition,
\begin{eqnarray}
{\cal B} (B^+ \rightarrow \pi^+ D_2^*(2460)^0) &<&
1.3 \times 10^{-3},   \nonumber  \\
{\cal B} (B^0 \rightarrow \pi^+ D_2^*(2460)^-) &<&
2.2 \times 10^{-3},   \nonumber  \\
{\cal B} (B^{+} \rightarrow \rho^+ D_2^*(2460)^{0}) &<& 4.7 \times
10^{-3},   \nonumber
\\ {\cal B} (B^{0} \rightarrow \rho^+ D_2^*(2460)^{-}) &<&
4.9 \times 10^{-3}~.
\label{expdata}
\end{eqnarray}
Recently the process $B \to K_2^* \gamma$ has been observed
for the first time by the CLEO Collaboration with a branching
ratio of $(1.66^{+0.59}_{-0.53} \pm 0.13) \times 10^{-5}$
\cite{cleo}, and by the Belle Collaboration with
${\cal B}(B\to K_2^*\gamma)=(1.50^{+0.58 +0.11}_{-0.53 -0.13}) \times 10^{-5}$
\cite{belle}.

Two-body hadronic $B$ decays involving a tensor meson $T$
($J^P = 2^+$) in the final state have long been studied
\cite{kv,cm,mrc,btopt,btovt}
using the non-relativistic quark model of Isgur, Scora, Grinstein
and Wise (ISGW) \cite{isgw} with the factorization
ansatz.  Some of those works \cite{kv,cm,mrc} studied $B$ decays involving
a $b \rightarrow c$ transition, which include the Cabibbo-Kobayashi-Maskawa
(CKM)-favored $B$ decays and the CKM-suppressed $B$ decays.
The estimated branching ratios of those decay modes strongly
depend on the properties of hadronic form factors. % calculated in the ISGW model.
A characteristic feature of the form factors given in the original ISGW
model \cite{isgw} is that values of the form factors decrease exponentially
as a function of $(t_m - t)$, where $t \equiv (p_B -p_T)^2$ is the
momentum transfer and $t_m \equiv (m_B -m_T)^2$ is the maximum possible momentum
transfer in the $B$ meson rest frame for a $B \to T$ transition. %in the model.
The authors in Ref. \cite{kv} used the form factors calculated at
the maximum momentum transfer $t_m$ for allowed transitions,
assuming that in the relevant transitions the momentum transfer
$(t)$ is close to the maximum momentum transfer $(t_m)$.  In
contrast, other authors \cite{mrc,btopt,btovt} used the form
factors with their exponentially decreasing behavior as a function
of $(t_m -t)$. In particular, in our previous works
\cite{btopt,btovt}, the exponentially decreasing behavior of the
form factors was assumed to predict the BR's of charmless decays
$B \to PT$ and $B \to VT$ ($P$ and $V$ denote a pseudoscalar and a
vector meson, respectively)\footnote{Recently the Belle
Collaboration measured the BR of $B^+ \to K^+ \pi^+ \pi^-$, where
two known candidate states for a $\pi^+ \pi^-$ invariant mass
around 1300 MeV are $f_2 (1270)$ and $f_0 (1370)$ \cite{bellePT}.
Because our previous result using the ISGW model predicts a rather
small BR for $B^+ \to f_2(1270) K^+$ \cite{btopt}, they concluded
that the measurements would provide evidence for a significant
nonfactorizable effect, if the peak were due to $f_2 (1270)$.
However, our recent result using the improved version of the model
(ISGW2) shows that the BR of $B^+ \to f_2(1270) K^+$ is enhanced
by an order of magnitude \cite{newbtoptvt}.}.

Because the exponentially decreasing behavior of the form factors
in the ISGW model is less justified, and the assumption of $t_m
\approx t$ seems to be too naive, it is very important to
carefully study all the relevant processes using more
\emph{reliable} and \emph{consistent} values of the form factors,
if available. In fact, the ISGW model has been improved to the
ISGW2 model \cite{isgw2}, whose feature includes a more accurate
parametrization of the form factors which have a more realistic
behavior at large $(t_m -t)$ by making the replacement of the
exponentially decreasing term to a certain polynomial term.  The
improved ISGW2 model also incorporates more reliable features, say
the constraints of heavy quark symmetry, relativistic corrections,
hyperfine distortions of wave functions, and so forth
\cite{isgw2}.

In this work we re-analyze $B \to PT$ and $B \to VT$ decays through a
$b \to c$ transition\footnote{In addition, we also study a few $B \to PT$ and $VT$ modes
involving a $b \to u$ transition, such as $B \to D_s^{(*)} a_2$,
$B \to D_s^{(*)} f_2^{(\prime)}$, and $B \to D^{(*)} K_2^*$.},
using the hadronic form factors calculated in the ISGW2 model.
We first discuss possible large uncertainties to the BR's of the relevant
modes, mainly arising from uncertainties to the hadronic
form factors which are heavily model-dependent.  Then, using the form factors
obtained in the ISGW2 model, we calculate the BR's, ratios of
${\cal B}(B\to VT)/{\cal B}(B\to PT)$ and CP asymmetries for $B \to PT$
and $B \to VT$.  We make comments on the difference between our results
and the previous results obtained using the relevant form
factors calculated in the original ISGW model.

This work is organized as follows.
In Sec. II we discuss uncertainties relevant to the hadronic form factors.
Our framework is introduced in Sec. III. We present our analysis of
$B \to PT$ and $B \to VT$ decays in Sec. IV.
Finally, our results are summarized in Sec. V.

%%%%%%%%%%%%%%%%%%%%%%%%%%%%%%%%%%%%%%%%%%%%%%%%%%%%%%%%%%%
%%%%%%%%%%%%%%%%%%%%%%%%%%%%%%%%%%%%%%%%%%%%%%%%%%%%%%%%%%%
\section{Uncertainties relevant to hadronic form factors}

The decay rate $(\Gamma)$ of $B \to PT$ or $B \to VT$ strongly depends on
the relevant hadronic form factors for $B \to T$ transitions.
%since $\Gamma \propto ({\rm form~ factors})^2$.
For instance, in $B \to PT$ decays, the decay rate is
\begin{eqnarray}
\Gamma (B \to PT) \propto (F^{B \to T})^2 ~,
\end{eqnarray}
where
\begin{eqnarray}
F^{B \to T} = k +(m_B^2 -m_T^2) b_+ +m_P^2 b_-~.
\end{eqnarray}
(See the next section for definitions of the form factors $k$, $b_+$, and $b_-$.)
Table I shows the values of the form factors $F^{B \to T}$ calculated in three cases:
(i) at $q^2 = m_D^2$ $(q^{\mu} \equiv p_B^{\mu} -p_T^{\mu})$,
(ii) at the maximum momentum transfer $t_m \equiv (m_B -m_T)^2$
in the ISGW model, and (iii)
at $q^2 = m_D^2$ in the ISGW2 model.
We note that $|F^{B \to T}| \approx 0.2$ at $t_m$,
while $|F^{B \to T}| \approx 0.05$ at $q^2 = m_D^2$ in the ISGW
model, where $T = a_2,~ f_2,~ f^{\prime}_2$.
The value of $|F^{B \to T}|$ calculated at $t_m$ is about 4 times larger than
that calculated at $q^2 = m_D^2$.
Thus, the decay rate of a relevant process
(e.g., $B \to D a_2$, $D f_2$, $D f_2^{\prime}$, etc)
evaluated by using the former value of the form factor (evaluated at $t_m$)
would be roughly 16 times
larger than that obtained using the latter value of the form factor
(at $q^2 = m_D^2$).
For $B \to VT$ decays, the similar argument holds.
It is obvious that the uncertainty relevant to the hadronic form factors can
seriously spoil theoretical estimates of the BR's of $B \to PT$ and $B \to VT$
decays.  More reliable values of the form factors are definitely needed.

As previously mentioned, a crucial improvement of the ISGW2 model is that
the form factors in this model have a more realistic and reasonable behavior
at large $(t_m -t)$.  Thus, one no longer needs to \emph{naively assume}
$t \approx t_m$ in $B \to PT$ and $VT$ processes.
The value of $|F^{B \to T}|$ obtained at $q^2 = m_D^2$
in the ISGW2 model is in between that obtained at $q^2 = m_D^2$
and that calculated at $t_m$ in the ISGW model (except $|F^{B \to K^*}|$).
In fact, from Table I, we see that for $B \to a_2$ and $B \to f_2$ transitions,
$|F^{B \to T}|$ obtained at $t_m$ is about 2 times larger than that obtained at
$q^2 = m_D^2$ in the ISGW2 model, which would lead to overestimation of the
relevant decay rates.
Compared to $|F^{B \to T}|$ obtained at $q^2 = m_D^2$
in the ISGW model, the values obtained in the ISGW2 model are about $2 - 6$
times larger, which would result in roughly $4 - 36$ times larger decay rates.

%%%%%%%%%%%%%%%%%%%%%%%%%%%%%%%%%%%%%%%%%%%%%%%%%%%%%%%%%%%
%%%%%%%%%%%%%%%%%%%%%%%%%%%%%%%%%%%%%%%%%%%%%%%%%%%%%%%%%%%
\section{Framework}

The relevant $\Delta B =1$ effective Hamiltonian for hadronic $B$ decays
can be written as
\begin{eqnarray}
H_{eff}^{q} &=& {G_F \over \sqrt{2}} \left[ V_{ub}V^{*}_{uq} (c_1 O^q_{1u}
+c_2 O^q_{2u}) + V_{cb}V^{*}_{cq} (c_1 O^q_{1c} +c_2 O^q_{2c}) \right.
\nonumber \\
&-& \left. \sum_{i=3}^{10} \left( V_{ub} V^*_{uq} c_i^u +V_{cb} V^*_{cq} c_i^c
+V_{tb} V^*_{tq} c_i^t \right) O_i^q \right]  \nonumber \\
&+& H.C.~ ,
\end{eqnarray}
where $O^q_i$'s are defined as
\begin{eqnarray}
O^q_{1f} &=& \bar q \gamma_{\mu} L f \bar f \gamma^{\mu} L b,  \ \
O^q_{2f} = \bar q_{\alpha} \gamma_{\mu} L f_{\beta} \bar f_{\beta} \gamma^{\mu}
L b_{\alpha}~,   \nonumber \\
O^q_{3(5)} &=& \bar q \gamma_{\mu} L b \sum_{q^{\prime}} \bar q^{\prime}
\gamma^{\mu} L(R) q^{\prime},  \ \
O^q_{4(6)} = \bar  q_{\alpha} \gamma_{\mu} L b_{\beta} \sum_{q^{\prime}}
\bar q^{\prime}_{\beta}  \gamma^{\mu}
L(R) q^{\prime}_{\alpha}~,  \nonumber \\
O^q_{7(9)} &=& {3 \over 2} \bar q \gamma_{\mu} L b \sum_{q^{\prime}}
e_{q^{\prime}} \bar q^{\prime}  \gamma^{\mu}
R(L) q^{\prime} , \ \
O^q_{8(10)} ={3 \over 2} \bar q_{\alpha} \gamma_{\mu} L b_{\beta}
\sum_{q^{\prime}}  e_{q^{\prime}} \bar
q^{\prime}_{\beta} \gamma^{\mu} R(L) q^{\prime}_{\alpha}~ ,
\end{eqnarray}
where $L(R) = (1 \mp \gamma_5)$, $f$ can be $u$ or $c$  quark, $q$
can be $d$ or $s$ quark,  and $q^{\prime}$ is summed over $u$,
$d$, $s$, and $c$ quarks. $\alpha$ and  $\beta$ are  the color
indices.  $T^a$ is the SU(3) generator with the normalization
${\rm Tr}(T^a T^b) =  \delta^{ab}/2$. $G^{\mu \nu}_a$ and $F^{\mu
\nu}$ are the gluon and photon field strength, and $c_i$'s  are
the Wilson  coefficients (WC's). We use the improved effective
WC's given in Ref.\cite{improvedwc}, where the renormalization
scheme- and scale-dependence of the WC's are discussed and
resolved.  The regularization scale is taken to be $\mu=m_b$
\cite{ddo}. The operators $O_1$, $O_2$ are the tree level and QCD
corrected operators, $O_{3-6}$ are the gluon induced strong
penguin operators, and finally  $O_{7-10}$ are the electroweak
penguin operators due to $\gamma$ and $Z$ exchange, and  the box
diagrams at loop level.

We use the improved ISGW2 quark model to analyze two-body nonleptonic decay
processes $B \to PT$ and $VT$ in the framework of generalized
factorization.  We describe the parameterizations of the hadronic
matrix elements in $B \to PT$ and $VT$ decays: \cite{isgw,isgw2}
\begin{eqnarray}
\langle 0 | A^{\mu} | P \rangle &=& i f_P p_P^{\mu} ~,  \\
\langle 0 | V^{\mu} | V \rangle &=& f_{_V} m_{_V} \epsilon^{\mu} ~,  \\
\langle T | j^{\mu} | B \rangle &=& i h(m_P^2) \epsilon^{\mu \nu
\rho \sigma} \epsilon^*_{\nu \alpha} p_B^{\alpha} (p_B
+p_T)_{\rho} (p_B -p_T)_{\sigma} + k(m_P^2) \epsilon^{* \mu \nu}
(p_B)_{\nu}  \nonumber \\
&\mbox{}&  + \epsilon^*_{\alpha \beta} p_B^{\alpha} p_B^{\beta} [
b_+(m_P^2) (p_B +p_T)^{\mu} +b_-(m_P^2) (p_B -p_T)^{\mu} ]~,
\label{formfactor}
\end{eqnarray}
where $j^{\mu} = V^{\mu} -A^{\mu}$.  $V^{\mu}$ and $A^{\mu}$
denote a vector and an axial-vector current, respectively.
$f_P~(f_V)$ denotes the decay constant of the relevant
pseudoscalar (vector) meson. $h(m_{P(V)}^2)$, $k(m_{P(V)}^2)$,
$b_+(m_{P(V)}^2)$, and $b_-(m_{P(V)}^2)$ express the form factors
for the $B \rightarrow T$ transition, which have been calculated
at $q^2 =m_{P(V)}^2$ $(q^{\mu} \equiv p_B^{\mu} -p_T^{\mu})$ in
the ISGW2 quark model. $p_B$ and $p_T$ denote the momentum of the
$B$ meson and the tensor meson, respectively. Using the above
parameterizations, the decay amplitudes for $B \to PT$ and $B \to
VT$ are \cite{btopt,btovt,Oh}
\begin{equation}
{\cal A}(B \to PT) \sim F^{B \to T}(m_P^2)~,~~~~~
{\cal A}(B\to VT) \sim \epsilon^{*\alpha\beta} F^{B \to T}_{\alpha\beta}(m_V^2)~,
\end{equation}
where
\begin{eqnarray}
F^{B \to T}(m_P^2) &=& k(m_P^2)+(m_B^2-m_T^2)b_+(m_P^2)+m_P^2b_-(m_P^2)~,
\label{FBT} \\
F^{B \to T}_{\alpha\beta}(m_V^2) &=& \epsilon^*_\mu(p_B+p_T)_\rho
 \Big[ih(m_V^2)\cdot \epsilon^{\mu\nu\rho\sigma}
  g_{\alpha\nu}(p_V)_\beta(p_V)_\sigma
 +k(m_V^2)\cdot\delta^\mu_\alpha\delta^\rho_\beta\nonumber\\
&& +b_+(m_V^2) \cdot(p_V)_\alpha(p_V)_\beta g^{\mu\rho}\Big]~.
\label{FBTv}
\end{eqnarray}
For our numerical analysis, we use the following values of the
decay constants (in MeV) \cite{sv}: $f_{\pi} = 132$, $f_{K} =
162$, $f_{D} = 252$, $f_{D_s} = 280$, $f_{\eta_c} = 393$,
$f_{\rho} = 216$, $f_{K^*} = 222$, $f_{D^*} = 249$, $f_{D_s^*} =
270$, $f_{J/\psi} = 405$. The running quark masses (in MeV) at
$m_b$ scale are used as follows \cite{do}: $m_u =3.6$, $m_d=6.6$,
and $m_s=100$.

An important feature of the ISGW2 model is that a more accurate
parametrization of the form factors $h$, $k$, $b_+$, and $b_-$ is
adopted by making the replacement, for $B \to T$ transition,
\begin{eqnarray}
{\rm exp}[- ({\rm constant}) \cdot (t_m -t)]~ \Rightarrow ~[1 + ({\rm constant})
\cdot (t_m -t)]^{-3}~,
\end{eqnarray}
where $t \equiv (p_B -p_T)^2$ is the
momentum transfer and $t_m \equiv (m_B -m_T)^2$ is the maximum possible momentum
transfer in the $B$ meson rest frame.
As a result, the form factors have a more realistic behavior at large
$(t_m -t)$.

We note that the matrix element $\langle 0 | j^{\mu} | T \rangle$
vanishes:
\begin{equation}
\langle 0 | j^{\mu} | T \rangle = p_\nu \epsilon^{\mu \nu}
(p_{_T}, \lambda) + p_{_T}^\mu \epsilon^\nu_{~\nu} (p_{_T},
\lambda) =0~,
\label{fT}
\end{equation}
because the trace of the polarization tensor $\epsilon^{\mu \nu}$
of the tensor meson $T$ vanishes and the auxiliary condition
holds, $p_T^{\mu} \epsilon_{\mu \nu} =0$ \cite{epsilon}.
Thus, in the generalized factorization scheme, any decay
amplitude for $B \rightarrow PT$ (or $VT$) is simply proportional to the
decay constant $f_P$ (or $f_V$) and a certain linear combination of
the form factors $F^{B \rightarrow T}$ (or $F^{B \to T}_{\alpha \beta}$),
{\it i.e.}, there is no such
amplitude proportional to $f_T \times F^{B \rightarrow P}$
(or $F^{B \to T}_{\alpha \beta}$) (see Appendix).

%%%%%%%%%%%%%%%%%%%%%%%%%%%%%%%%%%%%%%%%%%%%%%%%%%%%%%%%%%%%%%%%
%%%%%%%%%%%%%%%%%%%%%%%%%%%%%%%%%%%%%%%%%%%%%%%%%%%%%%%%%%%%%%%%
\section{Analyses and Results}

We calculate the BR's of $B \to PT$ and $B \to VT$ decays, whose quark
level processes are the $b \to c$ transition.
Among the relevant decay modes, many processes involve a tree diagram
only; their decay amplitudes are proportional to the CKM elements
$V_{cb}^* V_{ud}$ or $V_{cb}^* V_{us}$ (see Tables II$-$V).
Other processes involve both tree and (strong and electroweak) penguin
diagrams; their tree amplitudes are proportional to the CKM elements
$V_{cb}^* V_{cd}$ or $V_{cb}^* V_{cs}$.  But, the penguin diagram
contribution is much smaller than the tree contribution.
Expressions for all the amplitudes having both the tree and penguin
terms are presented in Appendix,
calculated in the generalized factorization scheme.
(For expressions of the other amplitudes, see Refs. \cite{kv,cm,mrc}.)
Among the relevant modes, some processes, such as
$B^{+ (0)} \to \pi^+ \bar D_2^{*0(-)}$,
$B^{+ (0)} \to D_s^+ \bar D_2^{*0(-)}$, etc, are the CKM-favored
decays whose amplitudes are proportional to the CKM elements
$V_{cb}^* V_{ud}$ (or $V_{cb}^* V_{cs}$ at tree level).
Other processes, such as $B^{+ (0)} \to K^+ \bar D_2^{*0(-)}$,
$B^{+ (0)} \to D^+ \bar D_2^{*0(-)}$, etc., are the CKM-suppressed
decays whose amplitudes are proportional to $V_{cb}^* V_{us}$
(or $V_{cb}^* V_{cd}$ at tree level).
As commented in the footnote of Sec. I, we also calculate
the BR's of some CKM-suppressed processes involving the $b \to u$
transition, such as $B^{+(0)} \to D_s^+ a_2^{0(-)}$,
$B^{+(0)} \to D^0 K_2^{*+(0)}$, and so on; these processes involve
a tree diagram only and their amplitudes are proportional to
$V_{ub}^* V_{cs}$.

Tables II and III show the BR's of $B \to PT$ processes for
$\Delta S =0$ and $|\Delta S| =1$, respectively ($S$ denotes the
strangeness quantum number).  Similarly, Tables IV and V show the
BR's of $B \to VT$ for $\Delta S =0$ and $|\Delta S| =1$,
respectively. In the tables, the results are shown for three
different values of the parameter $\xi \equiv 1/ N_c$ ($N_c$
denotes the effective number of color)\footnote{In the frameworks
of the QCD factorization and the perturbative QCD approaches, 
nonfactorizable
effects vary for different four-quark operators: e.g., $\xi$ is
different for tree- and penguin-dominated processes. But within our
generalized factorization framework, 
the $\xi$ is assumed universal. }: $\xi =0.1,~ 0.3,~
0.5$~.  For comparison, the BR's are also calculated using $a_1 =
1.15$ and $a_2 =0.26$ whose values are obtained from a fit to $B
\to PP$ and $B \to PV$ data \cite{a1a2}, where the QCD
coefficients are $a_1 \equiv c_1 + \xi c_2$ and $a_2 \equiv c_2
+\xi c_1$ ($c_1$ and $c_2$ are the effective WC's).

The decay amplitudes of all the modes shown in Tables II$-$V are
(dominantly) proportional to either $a_1$ (color-favored) or $a_2$
(color-suppressed) only. The value of $a_1 \equiv c_1 + \xi c_2$
does not vary much as $\xi$ varies: $a_1= 1.132$ for $\xi =0.1$,
$a_1= 1.059$ for $\xi =0.3$, and $a_1= 0.986$ for $\xi =0.5$. In
contrast, the value of $a_2 \equiv c_2 +\xi c_1$ varies as
follows: $a_2= -0.248$ for $\xi =0.1$, $a_2= -0.015$ for $\xi
=0.3$, and $a_2= 0.219$ for $\xi =0.5$.  We note that the value of
$a_2$ for $\xi =0.3$ is about an order of magnitude smaller than
that for $\xi =0.1$ or $\xi =0.5$. It would lead to the estimation
that the BR's of the decay modes, whose amplitudes are
proportional to $a_2$, are very small (i.e., about two orders of
magnitude smaller) for $\xi =0.3$. However, compared with the
values of $a_1$ and $a_2$ obtained from $B \to PP$ and $B \to PV$
data, the value of $a_2$ for $\xi= 0.3$ seems to be too small,
while the values of $a_1$ and $a_2$ for $\xi= 0.5$ are quite
consistent with those values. (For $\xi= 0.1$, the value of $a_1$
fits well to that obtained from the data, but $a_2$ has the
opposite sign to that deduced from the data. However, the sign of
$a_2$ has no (or negligible) effect on our results since each
decay amplitude is (dominantly) proportional to only one QCD
coefficient (i.e., either $a_1$ or $a_2$)).

As expected, the BR's of both the CKM-favored and color-favored processes
are generally large.
In $B \to PT$ decays, the BR of $B^{+(0)} \to \pi^+ \bar D_2^{*0(-)}$ is about
$3 \times 10^{-4}$, and the BR of $B^{+(0)} \to D^+_s \bar D_2^{*0(-)}$ is
about $4 \times 10^{-4}$.   In $B \to VT$ decays, the BR of
$B^{+(0)} \to \rho^+ \bar D_2^{*0(-)}$ is $(7-9) \times 10^{-4}$, and the BR of
$B^{+(0)} \to D^{*+}_s \bar D_2^{*0(-)}$ is about $1 \times 10^{-3}$.
The BR's of the CKM-favored and color-suppressed modes are $O(10^{-4}) - O(10^{-5})$,
except
${\cal B}(B^0 \to \bar D^0 f_2^{\prime}) \sim O(10^{-7})$ and
${\cal B}(B^0 \to \bar D^{*0} f_2^{\prime}) \sim O(10^{-6})$.
The BR's of the CKM-suppressed modes are relatively smaller,
$O(10^{-5}) - O(10^{-8})$.
{}From Tables II$-$V, we see that the BR's of the decay modes such as
$B^{+(0)} \to \bar D^{(*)0} a_2^{+(0)}$,
$B^0 \to \bar D^{(*)0} f_2^{(\prime)}$, $B^{+(0)} \to \eta_c (J/\psi) a_2^{+(0)}$,
$B^0 \to \eta_c (J/\psi) f_2^{(\prime)}$, etc, for $\xi =0.3$ are about two orders
of magnitudes smaller than those for $\xi =0.1$ or $\xi =0.5$.
This occurs because the decay amplitudes of all those modes are (dominantly)
proportional to $a_2$ (see Appendix), as explained above.

We note that for many processes our predictions are larger than the BR's
given in Ref. \cite{mrc}.  In particular, for the processes whose amplitudes are
proportional to $V^*_{ub} V_{cs}$, our results are about an order of magnitude
larger than the BR's given in \cite{mrc}; for instance, for $B \to PT$
such as $B^{+(0)} \to D_s^+ a_2^{0(-)}$, $B^+ \to D_s^+ f_2^{(\prime)}$,
$B^{+(0)} \to D^0 K_2^{*+(0)}$, and for $B \to VT$ such as
$B^{+(0)} \to D_s^{*+} a_2^{0(-)}$, $B^+ \to D_s^{*+} f_2^{(\prime)}$,
$B^{+(0)} \to D^{*0} K_2^{*+(0)}$.

In Table VI, we show the ratios of ${\cal B}(B \to VT) / {\cal B}(B \to PT)$.
The ratios are roughly 3 for the processes which involve a tree diagram only and
whose amplitudes are proportional to $a_1$ (via the external $W$
emission); for instance,
${\cal B}(B^{+(0)}\to \rho^{+} \Dbar_2^{*0(-)})
/{\cal B}(B^{+(0)}\to \pi^{+} \Dbar_2^{*0(-)}) \approx 3$.
This is naively expected from the fact that massive vector particles have three
polarization states.  But, for the processes which involve both tree and penguin
diagrams, the ratios deviate from 3; e.g.,
${\cal B}(B^{+(0)}\to D^{*+} \Dbar_2^{*0(-)})
/{\cal B}(B^{+(0)}\to D^{+} \Dbar_2^{*0(-)}) \approx 2.3$.
For the processes whose amplitudes are proportional to $a_2$ (via the internal $W$
emission), the ratios are $\sim 1.6$, except the processes involving $J/\psi$
or $\eta_c$ in the final state; e.g.,
${\cal B}(B^{+(0)}\to \Dbar^{*0} a_2^{+(0)})
/{\cal B}(B^{+(0)}\to \Dbar^{0} a_2^{+(0)}) \approx 1.6$.
We note that for the processes involving $J/\psi$ or $\eta_c$ in the final state,
the ratios substantially vary as $\xi$ varies from 0.1 to 0.3 to 0.5.
This is because the penguin contribution to the decay amplitudes involving
$\eta_c$ differs from that to the amplitudes involving $J/\psi$; the penguin
effect to the former amplitudes is proportional to the combination of the QCD
coefficients $(a_3 -a_5 +a_7 -a_9)$, while the penguin effect to the latter
amplitudes is proportional to $(a_3 +a_5 +a_7 +a_9)$.
We also compute CP asymmetries ${\mathcal A}_{CP}$ for $B \to PT$ and $B \to VT$
decays, defined by
\begin{eqnarray}
{\mathcal A}_{CP} = {{\mathcal B}(B \rightarrow f)  -{\mathcal
B}(\bar B \rightarrow \bar f) \over {\mathcal B}(B \rightarrow f)
+{\mathcal B}(\bar B \rightarrow \bar f)}~,
\end{eqnarray}
where $f$ and $\bar f$ denote a generic final state and its
CP-conjugate state. Since in the relevant modes the tree
contribution is very much dominant compared to the penguin
contribution, the asymmetries are relatively small\footnote{In
addition to the strong phases, there can be other
possible sources for the strong phases: for example, in the QCD
factorization a large strong phase for the WC, $a_2$, can be
induced by hard gluon exchange between final meson states, and in
the perturbative QCD approach large absorptive parts can be
generated from the weak annihilation diagrams. But, because in the
relevant modes the effect from the tree is much larger than that
from the penguin, as just mentioned in the text, the resultant
asymmetries would remain relatively small. }. We note that for a
non-vanishing ${\mathcal A}_{CP}$ for a process and its
CP-conjugate process, there should exist both the weak phase and
the strong phase differences between their tree and penguin
amplitudes. Thus, ${\mathcal A}_{CP}$'s vanish for the processes
involving $V^*_{cb} V_{cs}$ and $V^*_{tb} V_{ts}$, since there is
no weak phase in their amplitudes; e.g., ${\mathcal A}_{CP}
(B^{+(0)} \to D_s^+ \bar D_2^{*0(-)}) = 0$. We present our result
in Table VII. The CP asymmetries are shown for different values of
$\xi$. For all the relevant modes, the CP asymmetries are expected
to be a few percent.

%%%%%%%%%%%%%%%%%%%%%%%%%%%%%%%%%%%%%%%%%%%%%%%%%%%%%%%%%%%%%%%%
%%%%%%%%%%%%%%%%%%%%%%%%%%%%%%%%%%%%%%%%%%%%%%%%%%%%%%%%%%%%%%%%
\section{Conclusions}

We have studied the decay modes $B \to PT$ and $B \to VT$ whose quark level
processes are the $b \to c$ transition.  Due to large uncertainties to
the relevant hadronic form factors which are model-dependent, the previously
estimated BR's could be spoiled by large uncertainties.
Using more \emph{reliable} and \emph{consistent} values of the form factors
given in the improved version (ISGW2) of ISGW model, we re-calculate
the BR's of all the relevant modes
and find that for many modes our results are much larger than those given
in the previous work using the ISGW model.

Our results show that the BR's of some processes are quite large: in
$B \to VT$, the BR's of $B^{+(0)} \to D^{*+}_s \bar D_2^{*0(-)}$ and
$B^{+(0)} \to \rho^+ \bar D_2^{*0(-)}$ are $\sim 10^{-3}$, and in
$B \to PT$, the BR's of $B^{+(0)} \to D^+_s \bar D_2^{*0(-)}$ and
$B^{+(0)} \to \pi^+ \bar D_2^{*0(-)}$ are $(3-4) \times 10^{-4}$.
(These results are roughly consistent with those obtained under
the naive assumption of $t \approx t_m$ in the ISGW model.)
The estimated BR's of $B^{+(0)} \to \pi^+ \bar D_2^{*0(-)}$ and
$B^{+(0)} \to \rho^+ \bar D_2^{*0(-)}$ are about
a factor of $(4 - 5)$ smaller than the present experimental upper bounds
shown in Eq. (\ref{expdata}), and so far there is no known experimental
data on the modes $B \to D^{(*)}_s \bar D_2^*$.
Observations of these processes in $B$ experiments such as Belle, BaBar,
BTeV and LHC-B will be crucial in testing the ISGW2 model as well as validity
of the factorization scheme.
\\

%%%%%%%%%%%%%%%%%%%%%%%%%%%%%%%%%%%%%%%%%%%%%%%%%%%%%%
%%%%%%%%%%%%%%%%%%%%%%%%%%%%%%%%%%%%%%%%%%%%%%%%%%%%%%
\centerline{\bf ACKNOWLEDGEMENTS}
\medskip

\noindent We thank G. Cvetic for careful reading of the manuscript and his
valuable comments.
The work of C.S.K. was supported in part by  CHEP-SRC
Program, Grant No. 20015-111-02-2 and Grant No.  
R02-2002-000-00168-0 from BRP  of
the KOSEF, and in part by  Grant No. 2001-042-D00022
of the KRF. The work of J.-P.L. was supported by the BK21 Program.
The work of S.O. was supported by the KRF
Grants, Project No. 2001-042-D00022.

%%%%%%%%%%%%%%%%%%%%%%%%%%%%%%%%%%%%%%%%%%%%%%%%%%%%%%
%%%%%%%%%%%%%%%%%%%%%%%%%%%%%%%%%%%%%%%%%%%%%%%%%%%%%%

%%%%%%%%%%%%%%%%%%%%%%%%%%%%%%%%%%%%%%%%%%%%%%%%%%%%%%%%%%%%%%%%
%%%%%%%%%%%%%%%%%%%%%%%%%%%%%%%%%%%%%%%%%%%%%%%%%%%%%%%%%%%%%%%%
\newpage
\begin{center}
{\bf APPENDIX}
\end{center}

In this Appendix, we present expressions for the decay
amplitudes of $B \rightarrow PT$ and $VT$ modes which have both
the tree and penguin contributions (shown in Tables II$-$V).
Below we use $F^{B \to T}$, $F^{B \to T}_{\alpha \beta}$
and $X_{q q^{\prime}}$, defined by Eqs. (\ref{FBT})
and (\ref{FBTv}), and
\begin{eqnarray}
 X_{q q^{\prime}} = {m_P^2 \over (m_b +m_{q^{\prime}}) (m_q
+m_{q^{\prime}})}~.
\end{eqnarray}

(1) $B \rightarrow PT$ ($\Delta S = 0$) decays.
\begin{eqnarray}
A(B^+ \rightarrow D^+ \bar D_2^{*0}) &=& i {G_F \over \sqrt{2}} f_{D^+}
\epsilon^*_{\mu \nu} p^{\mu}_B p^{\nu}_B F^{B \rightarrow \bar D_2^{*0}}
(m^2_{D^+}) \left\{ V^*_{cb} V_{cd} a_1 \right.
\nonumber   \\
&\mbox{}& \left. - V^*_{tb} V_{td} [a_4
+a_{10} -2 (a_6 +a_8) X_{dc} ] \right\},
\\
A(B^0 \rightarrow D^+ D_2^{*-}) &=& i {G_F \over \sqrt{2}} f_{D^+}
\epsilon^*_{\mu \nu} p^{\mu}_B p^{\nu}_B F^{B \rightarrow D_2^{*-}}
(m^2_{D^+}) \left\{ V^*_{cb} V_{cd} a_1 \right.
\nonumber   \\
&\mbox{}& \left. - V^*_{tb} V_{td} [a_4
+a_{10} -2 (a_6 +a_8) X_{dc} ] \right\},
\\
A(B^+ \rightarrow \eta_c a^+_2) &=& i {G_F \over \sqrt{2}}
f_{\eta_c} \epsilon^*_{\mu \nu}
p^{\mu}_B p^{\nu}_B F^{B \rightarrow a^+_2} (m^2_{\eta_c}) \left\{
V^*_{cb} V_{cd} a_2 - V^*_{tb} V_{td} (a_3 -a_5 +a_7 -a_9 ) \right\},
\\
A(B^0 \rightarrow \eta_c a^0_2) &=& i {G_F \over 2}
f_{\eta_c} \epsilon^*_{\mu \nu}
p^{\mu}_B p^{\nu}_B F^{B \rightarrow a^0_2} (m^2_{\eta_c}) \left\{
V^*_{cb} V_{cd} a_2 - V^*_{tb} V_{td} (a_3 -a_5 +a_7 -a_9 ) \right\},
\\
A(B^0 \rightarrow \eta_c f_2) &=& i {G_F \over 2} \cos \phi_T
f_{\eta_c} \epsilon^*_{\mu \nu}
p^{\mu}_B p^{\nu}_B F^{B \rightarrow f_2} (m^2_{\eta_c}) \left\{
V^*_{cb} V_{cd} a_2 \right.
\nonumber   \\
&\mbox{}& \left. - V^*_{tb} V_{td} (a_3 -a_5 +a_7 -a_9 ) \right\},
\\
A(B^0 \rightarrow \eta_c f_2^{\prime}) &=& i {G_F \over 2} \sin \phi_T
f_{\eta_c} \epsilon^*_{\mu \nu}
p^{\mu}_B p^{\nu}_B F^{B \rightarrow f_2^{\prime}} (m^2_{\eta_c}) \left\{
V^*_{cb} V_{cd} a_2 \right.
\nonumber   \\
&\mbox{}& \left. - V^*_{tb} V_{td} (a_3 -a_5 +a_7 -a_9 ) \right\}.
\end{eqnarray}

%%%%%%%%%%%%%%%%%%%%%%%%%%%%%%%%%%%%%%%%%%%%%%%%%%%%%%%%%%%%%%
(2) $B \rightarrow PT$ ($|\Delta S| = 1$) decays.
\begin{eqnarray}
A(B^+ \rightarrow D^+_s \bar D_2^{*0}) &=& i {G_F \over \sqrt{2}} f_{D^+_s}
\epsilon^*_{\mu \nu} p^{\mu}_B p^{\nu}_B F^{B \rightarrow \bar D_2^{*0}}
(m^2_{D^+_s}) \left\{ V^*_{cb} V_{cs} a_1 \right.
\nonumber   \\
&\mbox{}& \left. - V^*_{tb} V_{ts} [a_4
+a_{10} -2 (a_6 +a_8) X_{sc} ] \right\},
\\
A(B^0 \rightarrow D^+_s D_2^{*-}) &=& i {G_F \over \sqrt{2}} f_{D^+_s}
\epsilon^*_{\mu \nu} p^{\mu}_B p^{\nu}_B F^{B \rightarrow D_2^{*-}}
(m^2_{D^+_s}) \left\{ V^*_{cb} V_{cs} a_1 \right.
\nonumber   \\
&\mbox{}& \left. - V^*_{tb} V_{ts} [a_4
+a_{10} -2 (a_6 +a_8) X_{sc} ] \right\},
\\
A(B^+ \rightarrow \eta_c K_2^{*+}) &=& i {G_F \over \sqrt{2}}
f_{\eta_c} \epsilon^*_{\mu \nu}
p^{\mu}_B p^{\nu}_B F^{B \rightarrow K_2^{*+}} (m^2_{\eta_c}) \left\{
V^*_{cb} V_{cs} a_2 - V^*_{tb} V_{ts} (a_3 -a_5 +a_7 -a_9 ) \right\},
\\
A(B^0 \rightarrow \eta_c K_2^{*0}) &=& i {G_F \over \sqrt{2}}
f_{\eta_c} \epsilon^*_{\mu \nu}
p^{\mu}_B p^{\nu}_B F^{B \rightarrow K_2^{*0}} (m^2_{\eta_c}) \left\{
V^*_{cb} V_{cs} a_2 - V^*_{tb} V_{ts} (a_3 -a_5 +a_7 -a_9 ) \right\}.
\end{eqnarray}

%%%%%%%%%%%%%%%%%%%%%%%%%%%%%%%%%%%%%%%%%%%%%%%%%%%%%%%%%%%%%%%%%%
%%%%%%%%%%%%%%%%%%%%%%%%%%%%%%%%%%%%%%%%%%%%%%%%%%%%%%%%%%%%%%%%%%

(3) $B \rightarrow VT$ ($\Delta S = 0$) decays.
\begin{eqnarray}
A(B^+ \rightarrow D^{*+} \bar D_2^{*0}) &=& i {G_F \over \sqrt{2}}
( m_{D^{*+}} f_{D^{*+}} \epsilon^{* \alpha \beta} F_{\alpha \beta}^{B
\rightarrow \bar D_2^{*0}} (m^2_{D^{*+}})) \left\{ V^*_{cb} V_{cd} a_1
- V^*_{tb} V_{td} (a_4 + a_{10}) \right\},
\\
A(B^0 \rightarrow D^{*+} \bar D_2^{*-}) &=& i {G_F \over \sqrt{2}}
( m_{D^{*+}} f_{D^{*+}} \epsilon^{* \alpha \beta} F_{\alpha \beta}^{B
\rightarrow \bar D_2^{*-}} (m^2_{D^{*+}})) \left\{ V^*_{cb} V_{cd} a_1
- V^*_{tb} V_{td} (a_4 + a_{10}) \right\},
\\
A(B^+ \rightarrow J/\psi a^+_2) &=& i {G_F \over \sqrt{2}}
( m_{J/\psi} f_{J/\psi} \epsilon^{* \alpha \beta} F_{\alpha \beta}^{B
\rightarrow a^+_2} (m^2_{J/\psi})) \left\{
V^*_{cb} V_{cd} a_2 \right.
\nonumber   \\
&\mbox{}& \left. - V^*_{tb} V_{td} (a_3 +a_5 +a_7 +a_9 ) \right\},
\\
A(B^0 \rightarrow J/\psi a^0_2) &=& i {G_F \over 2}
( m_{J/\psi} f_{JJ/\psi} \epsilon^{* \alpha \beta} F_{\alpha \beta}^{B
\rightarrow a^0_2} (m^2_{J/\psi})) \left\{
V^*_{cb} V_{cd} a_2 \right.
\nonumber   \\
&\mbox{}& \left. - V^*_{tb} V_{td} (a_3 +a_5 +a_7 +a_9 ) \right\},
\\
A(B^0 \rightarrow J/\psi f_2) &=& i {G_F \over 2} \cos \phi_T
( m_{J/\psi} f_{J/\psi} \epsilon^{* \alpha \beta} F_{\alpha \beta}^{B
\rightarrow f_2} (m^2_{J/\psi})) \left\{
V^*_{cb} V_{cd} a_2 \right.
\nonumber   \\
&\mbox{}& \left. - V^*_{tb} V_{td} (a_3 +a_5 +a_7 +a_9 ) \right\},
\\
A(B^0 \rightarrow J/\psi f_2^{\prime}) &=& i {G_F \over 2} \sin \phi_T
( m_{J/\psi} f_{J/\psi} \epsilon^{* \alpha \beta} F_{\alpha \beta}^{B
\rightarrow f_2^{\prime}} (m^2_{J/\psi})) \left\{
V^*_{cb} V_{cd} a_2 \right.
\nonumber   \\
&\mbox{}& \left. - V^*_{tb} V_{td} (a_3 +a_5 +a_7 +a_9 ) \right\}.
\end{eqnarray}

%%%%%%%%%%%%%%%%%%%%%%%%%%%%%%%%%%%%%%%%%%%%%%%%%%%%%%%%%%%%%%
(4) $B \rightarrow VT$ ($|\Delta S| = 1$) decays.
\begin{eqnarray}
A(B^+ \rightarrow D^{*+}_s \bar D_2^{*0}) &=& i {G_F \over \sqrt{2}}
( m_{D^{*+}_s} f_{D^{*+}_s} \epsilon^{* \alpha \beta} F_{\alpha \beta}^{B
\rightarrow \bar D_2^{*0}} (m^2_{D^{*+}_s})) \left\{ V^*_{cb} V_{cs} a_1
- V^*_{tb} V_{ts} (a_4 + a_{10}) \right\},
\\
A(B^0 \rightarrow D^{*+}_s D_2^{*-}) &=& i {G_F \over \sqrt{2}}
( m_{D^{*+}_s} f_{D^{*+}_s} \epsilon^{* \alpha \beta} F_{\alpha \beta}^{B
\rightarrow \bar D_2^{*-}} (m^2_{D^{*+}_s})) \left\{ V^*_{cb} V_{cs} a_1
- V^*_{tb} V_{ts} (a_4 + a_{10}) \right\},
\\
A(B^+ \rightarrow J/\psi K_2^{*+}) &=& i {G_F \over \sqrt{2}}
( m_{J/\psi} f_{J/\psi} \epsilon^{* \alpha \beta} F_{\alpha \beta}^{B
\rightarrow K_2^{*+}} (m^2_{J/\psi})) \left\{
V^*_{cb} V_{cs} a_2 \right.
\nonumber   \\
&\mbox{}& \left. - V^*_{tb} V_{ts} (a_3 +a_5 +a_7 +a_9 ) \right\},
\\
A(B^0 \rightarrow J/\psi K_2^{*0}) &=& i {G_F \over \sqrt{2}}
( m_{J/\psi} f_{J/\psi} \epsilon^{* \alpha \beta} F_{\alpha \beta}^{B
\rightarrow K_2^{*0}} (m^2_{J/\psi})) \left\{
V^*_{cb} V_{cs} a_2 \right.
\nonumber   \\
&\mbox{}& \left. - V^*_{tb} V_{ts} (a_3 +a_5 +a_7 +a_9 ) \right\}.
\end{eqnarray}

%%%%%%%%%%%%%%%%%%%%%%%%%%%%%%%%%%%%%%%%%%%%%%%%%%%%%%%%%%%%%%%%%%
%%%%%%%%%%%%%%%%%%%%%%%%%%%%%%%%%%%%%%%%%%%%%%%%%%%%%%%%%%%%%%%%%%
\newpage
\begin{table}
\caption{Form factors for $B \to T$ transitions calculated at $q^2
= m_D^2$ $(q^{\mu} \equiv p_B^{\mu} -p_T^{\mu})$, at the maximum
momentum transfer $t_m \equiv (m_B -m_T)^2$  in the ISGW model,
and at $q^2 = m_D^2$ in the ISGW2 model, respectively.}
\smallskip
\begin{tabular}{c|ccc}
Form factor for $B \to T$ & ISGW($m_D^2$) & ISGW($t_m$) & ISGW2
\\ \hline
   $F^{B \to a_2}$ &$-0.046$ &$-0.203$ &0.101
\\ $F^{B \to f_2}$ &$-0.045$ &$-0.205$ &0.099
\\ $F^{B \to f_2^{\prime}}$ &$-0.052$ &$-0.191$ &0.134
\\ $F^{B \to K_2^*}$ &$-0.049$ &$-0.111$ &0.131
\\ $F^{B \to D_2^*}$ &$-0.060$ &0.378 &0.367
\end{tabular}
\end{table}
%%%%%%%%%%%%%%%%%%%%%%%%%%%%%%%%%%%%%%%%%%%%%%%%%%%%%%%%%%%

%%%%%%%%%%%%%%%%%%%%%%%%%%%%%%%%%%%%%%%%%%%%%%%%%%%%%%%%%%%%%%%%
\begin{table}
\caption{Branching ratios of $B\to PT$ with $\Delta S=0$ in units
of $10^{-6}$, calculated in the ISGW2 model. }
\begin{center}
\begin{tabular}{c|cccc}
Decay mode &~~~ $\xi=0.1$~~~ & $\xi=0.3$~~~ & $\xi=0.5$~~~ &
$a_1=1.15$, $a_2=0.26$ \\\hline
$\propto V_{cb}^*V_{ud}$ & & & &\\
$B^+\to\pi^+ \Dbar_2^{*0}$ & 339.63 & 297.22 & 257.64 & 350.83 \\
$B^+\to \Dbar^0 a_2^{+}$ & 92.82 & 0.32 & 72.27 & 101.86 \\
$B^0\to \pi^{+} D_2^{*-}$ & 318.96 & 279.13 & 241.96 & 329.48 \\
$B^0\to \Dbar^0 a_2^0$ & 43.55 & 0.15 & 33.91 & 47.79 \\
$B^0\to \Dbar^0 f_2$ & 48.56 & 0.17 & 37.81 & 53.29 \\
$B^0\to \Dbar^0 f_2'$ & 0.57 & 0.002 & 0.44 & 0.62 \\\hline %%
%%------------------------------------------------------------
%%
$\propto V_{cb}^*V_{cd}$ & & & &\\
$B^+\to D^{+} \Dbar_2^{*0}$ & 22.23 & 19.45 & 16.86 & 22.68 \\
$B^+\to \eta_c a_2^+$ & 4.17 & 0.004 & 3.73 & 4.89 \\
$B^0\to D^{+} D_2^{*-}$ & 20.87 & 18.27 & 15.83 & 21.30 \\
$B^0\to \eta_c a_2^0$ & 1.96 & 0.002 & 1.75 & 2.30 \\
$B^0\to \eta_c f_2$ & 2.27 & 0.002 & 2.03 & 2.67 \\
$B^0\to \eta_c f_2'$ & 0.019 & 0.00002 & 0.017 & 0.02
\end{tabular}
\end{center}
\end{table}

%%%%%%%%%%%%%%%%%%%%%%%%%%%%%%%%%%%%%%%%%%%%%%%%%%%%%%%%%%%%%%%
\begin{table}
\caption{Branching ratios of $B\to PT$ with $|\Delta S|=1$ in
units of $10^{-6}$, calculated in the ISGW2 model. }
\begin{center}
\begin{tabular}{c|cccc}
Decay mode &~~~ $\xi=0.1$~~~ & $\xi=0.3$~~~ & $\xi=0.5$~~~ &
$a_1=1.15$, $a_2=0.26$ \\\hline
$\propto V_{cb}^*V_{cs}$ & & & &\\
$B^+\to D_s^{+}\Dbar_2^{*0}$ & 493.04 & 431.56 & 373.61 & 493.04 \\
$B^+\to \eta_c K_2^{*+}$ & 81.19 & 0.042 & 88.71 & 105.39 \\
$B^0\to D_s^{+} D_2^{*-}$ & 462.95 & 405.22 & 350.81 & 462.95 \\
$B^0\to \eta_c K_2^{*0}$ & 74.46 & 0.038 & 81.35 & 96.64 \\\hline
%%
%%----------------------------------------------------------------
%%
$\propto V_{cb}^*V_{us}$ & & & & \\
$B^+\to K^{+} \Dbar_2^{*0}$ & 24.64 & 21.56 & 18.69 & 25.45 \\
$B^+\to \Dbar^{0} K_2^{*+}$ & 6.69 & 0.023 & 5.21 & 7.34 \\
$B^0\to K^{+} D_2^{*-}$  & 23.14 & 20.25 & 17.56 & 23.91 \\
$B^0\to \Dbar^{0} K_2^{*0}$ & 6.19 & 0.021 & 4.82 & 6.80 \\\hline
%%
%%--------------------------------------------------------------
%%
$\propto V_{ub}^*V_{cs}$ & & & & \\
$B^+\to D_s^{+} a_2^0$ & 9.14 & 8.00 & 6.93 & 9.44\\
$B^+\to D_s^{+} f_2$ & 10.20 & 8.96 & 7.74 & 10.54 \\
$B^+\to D_s^{+} f_2'$ & 0.12 & 0.10 & 0.09 & 0.12 \\
$B^+\to D^0 K_2^{*+}$ & 1.07 & 0.004 & 0.83 & 1.17 \\
$B^0\to D_s^+ a_2^-$ & 17.15 & 15.01 & 13.01 & 17.71 \\
$B^0\to D^{0} K_2^{*0}$ & 0.99 & 0.003 & 0.77 & 1.08
\end{tabular}
\end{center}
\end{table}
%%%%%%%%%%%%%%%%%%%%%%%%%%%%%%%%%%%%%%%%%%%%%%%%%%%%%%%%%%%%%%%%
%%%%%%%%%%%%%%%%%%%%%%%%%%%%%%%%%%%%%%%%%%%%%%%%%%%%%%%%%%%%%%%

\begin{table}
\caption{Branching ratios of $B\to VT$ with $\Delta S=0$ in units
of $10^{-6}$, calculated in the ISGW2 model. }
\begin{center}
\begin{tabular}{c|cccc}
Decay mode & ~~~$\xi=0.1$~~~ & $\xi=0.3$~~~ & $\xi=0.5$~~~ &
$a_1=1.15$, $a_2=0.26$ \\\hline
$\propto V_{cb}^*V_{ud}$ & & & & \\
$B^+\to \rho^{+} \Dbar_2^{*0}$ & 950.15 & 831.51 & 720.77 & 981.48 \\
$B^+\to \Dbar^{*0} a_2^+$ & 151.77 & 0.53 & 118.16 & 166.54\\
$B^0\to \rho^{+} D_2^{*-}$ & 892.23 & 780.82 & 676.83 & 921.64 \\
$B^0\to \Dbar^{*0} a_2^0$ & 71.20 & 0.25 & 55.44 & 78.14 \\
$B^0\to \Dbar^{*0} f_2$ & 76.82 & 0.27 & 59.81 & 84.30 \\
$B^0\to \Dbar^{*0} f_2'$ & 0.95 & 0.003 & 0.74 & 1.05 \\\hline %%
%%----------------------------------------------------------------
%%
$\propto V_{cb}^*V_{cd}$ & & & & \\
$B^+\to D^{*+} \Dbar_2^{*0}$ & 50.05 & 43.78 & 37.94 & 53.25 \\
$B^+\to J/\psi a_2^+$ & 14.21 & 0.059 & 10.78 & 16.41 \\
$B^0\to D^{*+} D_2^{*-}$ & 46.98 & 41.10 & 35.61 & 49.99 \\
$B^0\to J/\psi a_2^0$ & 6.67 & 0.028 & 5.60 & 7.70 \\
$B^0\to J/\psi f_2$ & 7.28 & 0.03 & 5.53 & 8.41 \\
$B^0\to J/\psi f_2'$ & 0.074 & 0.0003 & 0.056 & 0.09
\end{tabular}
\end{center}
\end{table}
%%%%%%%%%%%%%%%%%%%%%%%%%%%%%%%%%%%%%%%%%%%%%%%%%%%%%%%%
\begin{table}
\caption{Branching ratios of $B\to VT$ with $|\Delta S|=1$ in
units of $10^{-6}$, calculated in the ISGW2 model. }
\begin{center}
\begin{tabular}{c|cccc}
Decay mode & ~~~$\xi=0.1$~~~ & $\xi=0.3$~~~ & $\xi=0.5$~~~ &
$a_1=1.15$, $a_2=0.26$ \\\hline
$\propto V_{cb}^*V_{cs}$ & & & & \\
$B^+\to D_s^{*+} \Dbar_2^{*0}$ & 1080.17 & 944.61 & 818.12 & 1200.8 \\
$B^+\to J/\psi K_2^{*+}$ & 307.66 & 1.66 & 224.02 & 383.62 \\
$B^0\to D_s^{*+} D_2^{*-}$ & 1013.89 & 886.64 & 767.92 & 1127.12 \\
$B^0\to J/\psi K_2^{*0}$ & 284.10 & 1.53 & 206.87 & 354.25
\\\hline %%
%%----------------------------------------------------------
%%
$\propto V_{cb}^*V_{us}$ & & & & \\
$B^+\to K^{*+} \Dbar_2^{*0}$ & 50.64 & 44.31 & 38.41 & 52.31 \\
$B^+\to \Dbar^{*0} K_2^{*+}$ & 11.04 & 0.038 & 8.59 & 12.11 \\
$B^0\to K^{*+} D_2^{*-}$  & 47.55 & 41.61 & 36.07 & 49.12 \\
$B^0\to \Dbar^{*0} K_2^{*0}$ & 10.25 & 0.035 & 7.98 & 11.24
\\\hline %%
%%---------------------------------------------------------
%%
$\propto V_{ub}^*V_{cs}$ & & & & \\
$B^+\to D_s^{*+} a_2^0$ & 15.00 & 13.13 & 11.38 & 15.49 \\
$B^+\to D_s^{*+} f_2$ & 16.17 & 14.15 & 12.26 & 16.70 \\
$B^+\to D_s^{*+} f_2'$ & 0.20 & 0.17 & 0.15 & 0.21 \\
$B^+\to D^{*0} K_2^{*+}$ & 1.76 & 0.006 & 1.37 & 1.93 \\
$B^0\to D_s^{*+} a_2^-$ & 28.15 & 24.63 & 21.35 & 29.08 \\
$B^0\to D^{*0} K_2^{*0}$ & 1.63 & 0.006 &1.27 & 1.79
\end{tabular}
\end{center}
\end{table}

\begin{table}
\caption{Ratios of ${\cal B}(B\to VT)/{\cal B}(B\to PT)$}
\begin{tabular}{c|ccc}
Ratio & $\xi =0.1$ & $\xi =0.3$ & $\xi =0.5$ \\\hline
%%%%%%%%%%%%%%%%%%%%%%%%%  Delta S = 0

${\cal B}(B^{+(0)}\to \rho^{+} \Dbar_2^{*0(-)})
/{\cal B}(B^{+(0)}\to \pi^{+} \Dbar_2^{*0(-)})$ &2.80 &2.80 &2.80 \\
${\cal B}(B^{+(0)}\to \Dbar^{*0} a_2^{+(0)})
/{\cal B}(B^{+(0)}\to \Dbar^{0} a_2^{+(0)})$ &1.64 &1.64 &1.63 \\
${\cal B}(B^0\to \Dbar^{*0} f_2)/{\cal B}(B^0\to \Dbar^{0} f_2)$ &1.58 &1.58 &1.58 \\
${\cal B}(B^0\to \Dbar^{*0} f_2')/{\cal B}(B^0\to \Dbar^{0} f_2')$
&1.68 &1.68 &1.68 \\\hline

${\cal B}(B^{+(0)}\to D^{*+} \Dbar_2^{*0(-)})
/{\cal B}(B^{+(0)}\to D^{+} \Dbar_2^{*0(-)})$ &2.25 &2.25 &2.25 \\
${\cal B}(B^{+(0)}\to J/\psi a_2^{+(0)})
/{\cal B}(B^{+(0)}\to \eta_c a_2^{+(0)})$ &3.41 &14.34 &2.89 \\
${\cal B}(B^0\to J/\psi f_2)/{\cal B}(B^0\to \eta_c f_2)$ &3.21 &13.50 &2.72 \\
${\cal B}(B^0\to J/\psi f_2')/{\cal B}(B^0\to \eta_c f_2')$ &3.83
&16.11 &3.24 \\\hline

%%%%%%%%%%%%%%%%%%%%%%%%%  |Delta S| = 1

${\cal B}(B^{+(0)}\to D_s^{*+} \Dbar_2^{*0(-)})
/{\cal B}(B^{+(0)}\to D_s^{+} \Dbar_2^{*0(-)})$ &2.19 &2.19 &2.19 \\
${\cal B}(B^{+(0)}\to J/\psi K_2^{*+(0)}) /{\cal B}(B^{+(0)}\to
\eta_c K_2^{*+(0)})$ &3.79 &39.82 &2.53 \\\hline

${\cal B}(B^{+(0)}\to K^{*+} \Dbar_2^{*0(-)})
/{\cal B}(B^{+(0)}\to K^{+} \Dbar_2^{*0(-)})$ &2.06 &2.06 &2.06 \\
${\cal B}(B^{+(0)}\to \Dbar^{*0} K_2^{*+(0)}) /{\cal
B}(B^{+(0)}\to \Dbar^{0} K_2^{*+(0)})$ &1.65 &1.65 &1.65 \\\hline

${\cal B}(B^{+(0)}\to D_s^{*+} a_2^{0(-)})
/{\cal B}(B^{+(0)}\to D_s^+ a_2^{0(-)})$ &1.64 &1.64 &1.64 \\
${\cal B}(B^+\to D_s^{*+} f_2)/{\cal B}(B^+\to D_s^{+} f_2)$ &1.58 &1.58 &1.58 \\
${\cal B}(B^+\to D_s^{*+} f_2')/{\cal B}(B^+\to D_s^{+} f_2')$ &1.69 &1.69 &1.69 \\
${\cal B}(B^{+(0)}\to D^{*0} K_2^{*+(0)}) /{\cal B}(B^{+(0)}\to
D^{0} K_2^{*+(0)})$ &1.65 &1.65 &1.65
\end{tabular}
\end{table}
%%%%%%%%%%%%%%%%%%%%%%%%%%%%%%%%%%%%%%%%%%%%%%%%%%%%%%%%
%%
%%%%%%%%%%%%%%%%%%%%%%%%%%%%%%%%%%%%%%%%%%%%%%%%%%%%%%%%
\begin{table}
\caption{CP asymmetries for $B \to PT$ and $B \to VT$}
\begin{tabular}{c|cccccc}
Decay mode & $\xi =0.1$ & $\xi =0.3$ & $\xi =0.5$  \\\hline
%%%%%%%%%%%%%%%%%%%%%%%%%  Delta S = 0  ONLY
%=================== B to PT ===============================
$B^{+(0)} \to D^{+} \Dbar_2^{*0 (-)}$ &$0.001$ &$0.001$ &$0.001$ \\
%=================== B to VT =====================================
$B^{+(0)} \to D^{*+} \Dbar_2^{*0(-)}$ &$-0.004$ &$-0.004$ &$-0.004$ \\
$B^{+(0)} \to J/\psi a_2^{+(0)}$ &$-0.0082$ &$-0.0045$ &$-0.0087$ \\
$B^0 \to J/\psi f_2^{(\prime)}$ &$-0.0082$ &$-0.0045$ &$-0.0087$
\end{tabular}
\end{table}
%%%%%%%%%%%%%%%%%%%%%%%%%%%%%%%%%%%%%%%%%%%%%%%%%%%%%%%%%%%

\end{document}